\documentclass[twocolumn,showpacs,preprintnumbers,amsmath,amssymb,aps]{revtex4}
\usepackage{amssymb}
\usepackage{graphicx}
\usepackage{amsmath}
\usepackage{amscd}
\usepackage{balance}
\usepackage[all]{xy}

\usepackage{fancyhdr}
\begin{document}
\title{Floquet system, Bloch oscillation, and Stark ladder}
\author{Tao Ma and Shu-Min Li}
\affiliation{Department of Modern Physics, University of Science and
Technology of China, P.O.Box 4, Hefei, Anhui 230026, People's
Republic of China}\date{\today}
\begin{abstract}
We prove the multi-band Bloch oscillation and Stark ladder in the
$nk$ and site representation from the Floquet theorem. The proof is
also possible from the equivalence between the Floquet system,
Bloch oscillation,
and the rotator with spin.
We also exactly solve the periodically driven two level atom and two
band Bloch oscillation in terms of Heun function.
\end{abstract}
\pacs{02.30.Ik, 42.50.-p, 72.15.-v}

\maketitle \thispagestyle{fancy}

\section{\label{sec:level1}Introduction}
Periodically driven two level atom with Rabi oscillation effect is
one of the simplest quantum systems, yet it still has not been
exactly solved \cite{Rabi1937, Bloch1940, Autler1955, Shirley1965,
Sambe1973, Rau1998, Wu2007}. The semiclassical Hamiltonian is the
interaction between a time periodic field and a two level atom. It
is the approximation to the interaction between a quantized photon
field and a two level atom when the the photon number is large
\cite{Shirley1965}. Floquet system (FS) has a time periodic
Hamiltonian. Periodically driven two level atom is the simplest FS.

Bloch oscillation (BO) and Stark ladder (SL) are an old and
controversial problem \cite{Bloch1929, Zener1934, Houston1940,
Wannier1960, Wannier1962, Zak1968,Fukuyama1973, Zak1977,
Krieger1986, Zakcomment1988, KriegerReply1988, Emin1987, Nenciu1991,
Rotvig1995}. One band BO and SL are easy to be established, but as
shown by Zak \cite{Zak1968}, there is a paradox in the one band BO.
Other proofs of BO and SL in the multi-band cases or in the case of
the general BO Hamiltonian includes \cite{Fukuyama1973, Zak1977,
Krieger1986, Emin1987}, but one proof is general followed by some
comment and reply. Nenciu argued when taking into consideration the
inter-band hopping, SL does not exist any more \cite{Nenciu1991}. In
the paper, we establish the multi-band BO and SL from the Floquet
theorem. Even if our proof is not general enough to end the
controversy, at least, it points out the physical mechanism of BO
and SL: the Floquet theorem.

The BO Hamiltonian in the $nk$ representation allows an
interpretation as a FS. Based on the Floquet theorem, we give a
proof of BO and SL both in the $nk$ and site representation.

\section{\label{sec:level1}Exact solution of a periodically driven two level atom}
The Hamiltonian of a periodically driven two level atom is
\begin{equation}
H(t)=\Omega \sigma_z-A \sigma_x \sin\omega t,
\end{equation}
where $\sigma_x$ and $\sigma_z$ are Pauli matrices, and $\Omega$ and
$A$ are parameters. After a time-independent unitary rotation in the
Hilbert space \cite{Barata2000}, the Hamiltonian becomes
\begin{equation}
H'(t)=A \sigma_z \sin\omega t+\Omega \sigma_x.
\end{equation}
The Schrodinger equation is
\begin{equation}
\begin{split}
i\hbar\frac{\partial}{\partial t} \left(
\begin{array}{l}
 c_1 \\
 c_2
\end{array}
\right) &=H'(t) \left(
\begin{array}{l}
 c_1 \\
 c_2
\end{array}
\right) \\
&=(A \sigma_z \sin\omega t+\Omega \sigma_x) \left(
\begin{array}{l}
 c_1 \\
 c_2
 \end{array}
\right),
\end{split}
\end{equation}
where $(c_1,c_2)^T$ is the wave function of the two level atom after
the unitary transform. For simplicity, we set $\hbar=1$ and
$\omega=1$.
\begin{eqnarray}
(i\frac{\partial}{\partial t}-A \sin t)c_1&=&\Omega c_2, \nonumber\\
(i\frac{\partial}{\partial t}+A \sin t)c_2&=&\Omega c_1 .
\end{eqnarray}
After removing $c_2$, we get
\begin{equation}
\frac{\partial^2}{\partial t^2} c_1+(i A \cos t+A^2 \sin^{2} t
+\Omega^2 )c_1=0.
\end{equation}
We change the variable from $t$ to $z$. $z=\frac{1}{2}\cos
t+\frac{1}{2}$. We get
\begin{equation}
\begin{split}
&(z^2-z)\frac{\partial^2}{\partial z^2}c_1
+(z-\frac{1}{2})\frac{\partial}{\partial z}c_1 \\
&+(4A^2 z^2-4A^2 z-2 i A z+i A-\Omega^2)c_1=0.
\end{split}
\end{equation}
The general solution is
\begin{equation}
\begin{split}
&c_1(z)=e^{2iAz} \times \\
&\bigg\{
\begin{array}{l}
 H_c(4iA,-\frac{1}{2},-\frac{1}{2},-2iA,iA+\frac{3}{8}-\Omega^2,z) \\
 \sqrt{z}H_c(4iA,\frac{1}{2},-\frac{1}{2},-2iA,iA+\frac{3}{8}-\Omega^2,z)
\end{array} ,
\end{split}
\end{equation}
where $H_c$ is the Heun confluent function. See the appendix of the
Heun function. The general solution is
\begin{equation}
\begin{split}
&c_1(t)=e^{2iA(\cos t+1)} \times \\
&\bigg\{
\begin{array}{l}
 H_c(4iA,-\frac{1}{2},-\frac{1}{2},-2iA,iA+\frac{3}{8}-\Omega^2,\frac{1}{2}\cos t+\frac{1}{2}) \\
 H_c(4iA,\frac{1}{2},-\frac{1}{2},-2iA,iA+\frac{3}{8}-\Omega^2,\frac{1}{2}\cos
 t+\frac{1}{2}) \times \\ \quad(\cos t+1)^{1/2},
\end{array}
\end{split}
\end{equation}
When $t=0$, $z=1$, which is a singular point of $H_c$. Now we do not
have enough knowledge to handle the singular point in order to
consider the initial conditions. When
\begin{equation}
H'(t)=A \sigma_x \cos t+\Omega \sigma_z.
\end{equation}
The general solution is
\begin{equation}
\begin{split}
&c_1(t)=\bigg \{
\begin{array}{l}
 H_d(4A,-4\Omega^2,-8 A,4\Omega^2,i \cot\frac{t}{2})e^{-iA\sin t} \\
 H_d(-4A,-4\Omega^2,-8 A,4\Omega^2,i \cot\frac{t}{2})e^{iA\sin t}.
\end{array}
\end{split}
\end{equation}

\section{\label{sec:level1}Exact solution of two band Bloch oscillation}
The Hamiltonian of two band Bloch electron in a electric field is
\begin{equation}
\begin{split}
H=&\sum_{n=-\infty}^{\infty} [(\Delta-eF
n)a_{n}^{\dagger}a_{n}+(-\Delta-eF n)b_{n}^{\dagger}b_{n}\\
&+\frac{1}{2}t_a(a_{n+1}^{\dagger}a_{n}+a_{n}^{\dagger}a_{n+1})+
\frac{1}{2}t_b(b_{n+1}^{\dagger}b_{n}+b_{n}^{\dagger}b_{n+1})\\
&-eFR(a_{n}^{\dagger}b_{n}+b_{n}^{\dagger}a_{n})],
\end{split}
\end{equation}
where $a$ and $b$ refer to electrons in two bands with bandwidths
$2t_a$ and $2t_b$ respectively; the first two terms are the site
energies, the middle two describe site-to-site hopping, and the last
term gives interband hopping \cite{Fukuyama1973, Rotvig1995}.

The Hamiltonian can be Fourier transformed \cite{Fukuyama1973,
Hartmann2004} into the $nk$ representation \cite{Kane1959,
Callaway1963,Zak1968}, in which the Hamiltonian is
\begin{equation}
H=
\begin{bmatrix}
\Delta-ieF\frac{\partial}{\partial k}+t_a \cos k & -eFR
\\-eFR
&-\Delta-ieF\frac{\partial}{\partial k}+t_b \cos k &
\end{bmatrix}.
\end{equation}
The eigenstates of the Hamiltonian is periodic of $k$
\cite{Zak1968}. We assume the period is $2\pi$.

We define
\begin{equation}
H_R=-ieF\frac{\partial}{\partial k}+\begin{bmatrix} t_a \cos k &0
\\0
&t_b \cos k &
\end{bmatrix}
\end{equation} as the rotation Hamiltonian
and
\begin{equation}
H_S=
\begin{bmatrix}
\Delta & -eFR
\\-eFR
&-\Delta &
\end{bmatrix}
\end{equation}
as the spin Hamiltonian. The names will be explained in the Sec. IV.

\subsection{Rotation and spin decoupled}
Now we discuss the simplest two band BO. If $t_a=t_b=t$,
\begin{equation}
H=-ieF\frac{\partial}{\partial k}+t \cos k+
\begin{bmatrix}
\Delta & -eFR
\\-eFR
&-\Delta &
\end{bmatrix}.
\end{equation}
Then $[H_R,H_S]=0$. The rotation and spin degrees of freedom are
decoupled. The eigenvalues of Eq. $(15)$ is
\begin{equation}
\omega_{n\pm}=n eF \pm \sqrt{\Delta^2+(eFR)^2}.
\end{equation}
The eigenstates are
\begin{equation}
\phi_{n\pm}=e^{i n k-i t\sin k}\begin{bmatrix}-\frac{1}{R}(\Delta
\pm \sqrt{\Delta^2+(eFR)^2})\\1\end{bmatrix}
\end{equation}
The result is first derived by Fukuyama \textit{et al} in
\cite{Fukuyama1973}. The physical meaning of Eq. $(16,17)$ is two
SLs. Electron oscillates between two bands with a period $\frac{\pi
R}{\sqrt{\Delta^2+R^2}}$.

\subsection{Exact solution of two band Bloch oscillation}
When $[H_R, H_S] \neq 0$, the rotation and spin degrees of freedom
are coupled together. The Hamiltonian eigenvalue equation is
\begin{equation}
H\psi=E\psi.
\end{equation}

\begin{equation}
\begin{split}
&\begin{bmatrix} \Delta-ieF\frac{\partial}{\partial k}+t_a \cos k &
-eFR
\\-eFR
&-\Delta-ieF\frac{\partial}{\partial k}+t_b \cos k &
\end{bmatrix}
\left(
\begin{array}{l}
 c_1 \\
 c_2
\end{array}
\right) \\
&\qquad \qquad \qquad \qquad \qquad =E \left(
\begin{array}{l}
 c_1 \\
 c_2
\end{array}
\right),
\end{split}
\end{equation}
where $\psi=(c_1,c_2)^T$ is the wave function and $c_1$ and $c_2$
are functions of $k$.
\begin{eqnarray}
(-i e F \frac{\partial}{\partial k}+t_a \cos k+\Delta-E)c_1-e F R c_2&=&0, \nonumber\\
-e F R c_1+(-i e F \frac{\partial}{\partial k}+t_b \cos
k-\Delta-E)c_2&=&0.
\end{eqnarray}
We set the unit $eF=1$. After removing $c_2$,
\begin{equation}
\begin{split}
&\frac{\partial^2}{\partial k^2}c_1+\big( (t_a+t_b)\cos k-2E \big)i
\frac{\partial}{\partial
k}c_1 \\
&+\bigg(-i t_a \sin k- t_a t_b \cos^2 k-(t_a
(-\Delta-E)\\
&\qquad +t_b (\Delta-E))\cos k-E^2+\Delta^2 +R^2\bigg)c_1=0.
\end{split}
\end{equation}

We change the variable from $c_1(k)$ to $u(k)$ and define
$t_a-t_b=\delta$. $c_1(k)=u(k)\exp(-i a\sin k)$.

\begin{equation}
\begin{split}
&\frac{\partial^2}{\partial k^2}u-\big(\delta \cos k+2 E)i
\frac{\partial}{\partial k}u \\
&\quad \quad+\bigg(\delta(\Delta-E)\cos k+\Delta^2+R^2-E^2\bigg)u=0.
\end{split}
\end{equation}

Eq. $(22)$ can be solve by Heun function. What we have to do is to
transform it into the standard form of Heun function. We change
variable from $k$ to $z$. $z=I \cot\frac{k}{2}$.
\begin{widetext}
\begin{equation}
\begin{split}
&(z-1)^3 (z+1)^3 \frac{\partial^2}{\partial
z^2}u+[2z^5-(4E+2\delta)z^4-4z^3+8Ez^2+2z+(-4E+2\delta)]\frac{\partial}{\partial
z}u \\
&+[4(E^2-\Delta^2-R^2-\delta \Delta +\delta
E)z^2+4(-E^2+\Delta^2+R^2-\delta \Delta +\delta E)]u=0.
\end{split}
\end{equation}

After another variable change, Eq. $(23)$ can be written into the
standard form. The general solution is

\begin{equation}
u(z)=(\frac{-1+z}{1+z})^E\bigg\{
\begin{array}{l}
 H_d\left( 2\delta,-4\delta\Delta-4{\Delta}^{2}-4{R}
^{2},-4\delta,-4\delta\Delta+4{\Delta}^{2}+4{R}^{2},z
 \right)
 \\
H_d\left( -2\delta,-4\delta\Delta-4{\Delta}^{2}-4{R
}^{2},-4\delta,-4\delta\Delta+4{\Delta}^{2}+4{R}^{2},z
 \right) e^{-2\delta z/(z^2-1)}
\end{array} \bigg\}.
\end{equation}

\begin{equation}
u(k)=e^{iE k}\bigg\{
\begin{array}{l}
 H_d\left( 2\delta,-4\delta\Delta-4{\Delta}^{2}-4{R}
^{2},-4\delta,-4\delta\Delta+4{\Delta}^{2}+4{R}^{2},i\cot(\frac{k}{2})
 \right)
 \\
H_d\left( -2\delta,-4\delta\Delta-4{\Delta}^{2}-4{R
}^{2},-4\delta,-4\delta\Delta+4{\Delta}^{2}+4{R}^{2},i\cot(\frac{k}{2})
 \right) e^{i\delta\sin k}
\end{array} \bigg\}.
\end{equation}

\begin{equation}
c_1(k)=e^{iE k}\bigg\{
\begin{array}{l}
 H_d\left( 2\delta,-4\delta\Delta-4{\Delta}^{2}-4{R}
^{2},-4\delta,-4\delta\Delta+4{\Delta}^{2}+4{R}^{2},i\cot(\frac{k}{2})
 \right) e^{-i t_a\sin k}
 \\
H_d\left( -2\delta,-4\delta\Delta-4{\Delta}^{2}-4{R
}^{2},-4\delta,-4\delta\Delta+4{\Delta}^{2}+4{R}^{2},i\cot(\frac{k}{2})
 \right) e^{-i t_b\sin k}
\end{array} \bigg\}.
\end{equation}
\end{widetext}

\section{\label{sec:level1}Floquet system, Bloch oscillation, and Stark ladder}
The Schrodinger equation of a FS is
\begin{equation}
i\hbar\frac{\partial}{\partial t}\Psi(t)=H(t)\Psi(t),
\end{equation}
where $\Psi(t)$ is the wave function and $H(t)=H(t+2\pi)$. Due to
the Floquet theorem, the fundamental solutions of Eq. $(27)$ are the
multiple of a c-number and a time periodic wave function.
\begin{equation}
\Psi(t)=e^{-iEt/\hbar}\Phi,
\end{equation}
where $E$ is a real number and referred as quasienergy
\cite{Sambe1973}. It can be verified
\begin{equation}
(H(t)-i\hbar\frac{\partial}{\partial t})\Phi(t)=E\Phi(t).
\end{equation}
Now we change the variable from the time $t$ to the Bloch number
$k$. $t=k$.
\begin{equation}
(H(k)-i\hbar\frac{\partial}{\partial k})\Phi(k)=E\Phi(k).
\end{equation}
$\Phi(k)$ is seen as the wave function of the Bloch electron.

The general BO Hamilton
\begin{equation}
H=\frac{p^2}{2m}+V(x)-e F x,
\end{equation}
where $m, x, p$ are the mass, position and momentum of the Bloch
electron. In the $nk$ representation, a multi-band approximation of
Eq. $(31)$ is \cite{Kane1959, Callaway1963,Zak1968,Zak1977}
\begin{equation}
H(k)=\sum_{n=1}^M\epsilon_n(k)-\sum_{n,m=1}^M eF X_{nm}(k) -i
eF\frac{\partial}{\partial k},
\end{equation}
where $X_{nm}(k)$
\begin{equation}
X_{nm}(k)=\int u_{nk}^{\ast}(x)i\frac{\partial}{\partial k}u_{mk}(x)
\, dx
\end{equation}
is periodic of $k$.

Eq. $(30)$ and $(32)$ have the same form. The eigenvalue problem of
the BO Hamiltonian is
\begin{equation}
\bigg[\sum_{n=1}^M\epsilon_n(k)-\sum_{n,m=1}^M eF X_{nm}(k) -i
eF\frac{\partial}{\partial k} \bigg]\Phi(k)=E \Phi(k),
\end{equation}

If $\{E, \Phi(k)\}$ is the eigenvalue and eigenstate of Eq. $(30)$
and $(32)$, then
\begin{align}
&E'=E+m\times 2\pi E_e, \\
&\Phi'(k)=e^{i m \times 2\pi k}\Phi(k),
\end{align}
gives another eigenvalue and eigenstate of the BO Hamiltonian. But
it gives the same solution to the FS Eq. $(27)$. In FS, if $H(t)$ is
a $M \times M$ matrix, which corresponds to a $M$ band BO
Hamiltonian, the total number of the quasienergy is $M$. The
eigenvalues of the BO Hamiltonian are grouped into $M$ SLs. If $M$
is finite, BO can not have continuous spectra and the eigenstates of
BO Hamiltonian are localized. If the electron is put on one site,
the wave function oscillates because the wave function can be
expanded by (approximately) finite eigenstates of the BO
Hamiltonian. It is the connection of Eq. $(32)$ to the Floquet
system, that gives the eigenvalues of SLs. We note Fukuyama
\textit{et al} first used the Floquet theorem to prove SL in a two
band case \cite{Fukuyama1973}. Avron \textit{et al} also gave a
proof of SL based on Eq. $(32)$ in \cite{Zak1977}. But our proof is
clearer.

We require $H(k)$ is not a function of $i\frac{\partial}{\partial
k}$. In this way, we implicitly assume
\begin{equation}
\int u_{nk}^{\ast}(x)u_{mk}(x) \, dx = 0
\end{equation}
or
\begin{equation}
\langle ns|  ms \rangle=0
\end{equation}
when $n \neq m$ in the site representation. Under this assumption,
the interband hopping matrix elements $X_{nm}(k)$ or
\begin{equation}
\langle ns|-e F (x)|  ms \rangle=\langle ns|-e F (x+x_0)|  ms
\rangle
\end{equation}
do not depends on $x$.

\subsection{Infinite matrix representation}
Every FS and BO Hamiltonian can be represented as an infinite
matrix. The infinite matrix of the periodically driven two level
atom was first written out and referred as Floquet Hamiltonian by
Shirley \cite{Shirley1965}. The FS formalism \cite{Shirley1965,
Sambe1973} including the infinite matrix representation is widely
used in the practical numerical calculations, such as atomic and
molecular excitation, ionization in a laser field \cite{Chu2004}.
The Hamiltonian of a periodically driven two level atom is
\begin{equation}
H(t)=\Omega \sigma_z+A\cos (\omega t) \sigma_x.
\end{equation}
Then the Floquet Hamiltonian
\begin{equation}
\mathcal {H}_F=H(t)-i\frac{\partial}{\partial t},
\end{equation}
has the following infinite matrix representation
\cite{Shirley1965,Sambe1973,Chu2004} in the $|s n \rangle=|\uparrow,
\frac{1}{\sqrt{2\pi}}e^{-i n t} \rangle$ or $|\downarrow,
\frac{1}{\sqrt{2\pi}}e^{-i n t} \rangle$ basis \small
\begin{widetext}
\begin{equation}
   \left(
\begin{array}{llllllllll}
    \cdots & \cdots &   &  &   &   &   &   &   &   \\
    \cdots  & -\Omega +2\omega & A/2 & 0 & 0 & 0 & 0 & 0 & 0 &  \\
      & A/2 & \Omega +\omega  & 0 & 0 & A/2 & 0 & 0 & 0 &   \\
      & 0 & 0 & -\Omega+\omega & A/2 & 0 & 0 & 0 & 0 &   \\
      & 0 & 0 & A/2 & \Omega  & 0 & 0 & A/2 & 0 &  \\
      & 0 & A/2 & 0 & 0 & -\Omega  & A/2 & 0 & 0 &  \\
      & 0 & 0 & 0 & 0 & A/2 & \Omega-\omega   & 0 & 0 & \\
      & 0 & 0 & 0 & A/2 & 0 & 0 & -\Omega-\omega  & A/2 & \\
      & 0 & 0 & 0 & 0 & 0 & 0 & A/2 & \Omega-2\omega & \cdots \\
      &   &   &   &   &   &   &   & \cdots & \cdots
   \end{array}
   \right),
\end{equation}
\end{widetext}
\normalsize
which has a more compact form \small
\begin{equation}
\left(
\begin{array}{lllllll}
\cdots & \cdots &  &  &  &  &  \\
\cdots   & H_S+2 \omega  & H_R & 0 & 0 & 0 &  \\
   & H_R & H_S+\omega  & H_R & 0 & 0 &  \\
   & 0 & H_R & H_S & H_R & 0 &  \\
   & 0 & 0 & H_R & H_S-\omega  & H_R &  \\
   & 0 & 0 & 0 & H_R & H_S-2 \omega  & \cdots \\
   &   &   &   &   & \cdots & \cdots
\end{array}
\right),
\end{equation}
\normalsize
where
\begin{equation}
H_S=
\begin{bmatrix}
\Omega & 0 \\
0 &-\Omega &
\end{bmatrix};
H_R=
\begin{bmatrix}
0 & A/2 \\
A/2 &0 &
\end{bmatrix}.
\end{equation}
The infinite matrix representation of the two band BO Hamiltonian
Eq. $(11)$ is \small
\begin{equation}
\left(
\begin{array}{lllllll}
\cdots & \cdots &  &  &  &  &  \\
\cdots   & H_S+2 eF  & H_R & 0 & 0 & 0 &  \\
   & H_R & H_S+eF  & H_R & 0 & 0 &  \\
   & 0 & H_R & H_S & H_R & 0 &  \\
   & 0 & 0 & H_R & H_S-eF  & H_R &  \\
   & 0 & 0 & 0 & H_R & H_S-2 eF  & \cdots \\
   &   &   &   &   & \cdots & \cdots
\end{array}
\right),
\end{equation}
\normalsize

where
\begin{equation}
H_S=
\begin{bmatrix}
\Delta & -e F R \\
-e F R &-\Delta &
\end{bmatrix};
H_R=
\begin{bmatrix}
t_a/2 & 0 \\
0 &t_b/2 &
\end{bmatrix}.
\end{equation}
Note in this section $H_R$ is different from the Sec. III by the
part of $-ieF\frac{\partial}{\partial k}$.

The Floquet Hamiltonian of periodically driven two level atom Eq.
$(43,44)$ and the two band BO Hamiltonian Eq. $(45,46)$ are very
similar. The difference is just the different basis of the Hilbert
space and parameters. Given a FS, we can find a BO Hamiltonian,
which has the same infinite matrix representation as the Floquet
Hamiltonian of the FS and \textit{vice versa}. FS must has very
similar dynamic behaviors with BO and \textit{vice versa}. This is
the reason why we can use the Floquet theorem to prove BO.

\subsection{Floquet system, Bloch oscillation, and rotator with spin}
The Hamiltonian Eq. $(32)$ is an approximation of Eq. $(31)$ based
on the Bloch function. The proof of BO and SL is achieved in the
$nk$ representation. But Eq. $(31)$ can also be approximated in the
site representation just as two band approximation Eq. $(11)$.

The BO Hamiltonian of a multi-band Bloch electron in a linear
electric field is a rotator with the spin degree of freedom (RS).
The basis of the Hilbert space of Eq. $(11)$ is $|s n\rangle$, where
$s$ is one of the multi-bands and $n$ is $n$-th site. In the rotator
representation, the basis of the Hilbert space is $|s n\rangle$,
where $s$ is the spin of the rotator and $n$ is
$\frac{1}{\sqrt{2\pi}}e^{-i n k}$ with $k$ the angle of the rotator.
In the Sec. III, we rewrite Eq. $(11)$ as Eq. $(12)$ from the
perspective of the Fourier transform. But this can also be achieved
by the site-band and rotation-spin correspondence. The $n$-th site
corresponds to $\frac{1}{\sqrt{2\pi}}e^{-i n k}$ and the $s$-th band
the $s$-th spin state. From this correspondence, we can rewrite Eq.
$(11)$ into Eq. $(12)$ without the Fourier transform. Every BO
Hamiltonian can be rewritten as a RS. $H(k)$, such as in the Eq.
$(12)$, is the coupling between rotation-rotation, spin-spin and
rotation-spin degrees of freedom. The Bloch wave number in the $nk$
representation $k$ is the angle of the rotator. By corresponding the
Hilbert space basis and the Hamiltonian, we have established the
equivalence between BO and RS.

For the Hamiltonian of every one of FS, BO and RS, we can find
another two Hamiltonians in the other two systems, and the former
Hamiltonian has the equivalent behavior with the latter two.

Generalizing Eq. $(12)$, in the rotator representation, Eq. $(31)$
is approximated as
\begin{equation}
\begin{split}
H&=\sum_{s=1}^M\epsilon_s(k)|s\rangle \langle s|\\
&-\sum_{s,s'=1}^M \sum_{n,m=-\infty}^\infty \sum_{s n\neq s' m} eF
X_{s n,s' m}|s n\rangle\langle s' m| -i eF\frac{\partial}{\partial
k},
\end{split}
\end{equation}
where $|s n\rangle = |s,  \frac{1}{\sqrt{2\pi}}e^{-i n k} \rangle$
is the basis of the RS and
\begin{equation}
X_{s n,s' m}=\int u_{s n}^{\ast}(x)x u_{s' m}(x) \, dx
\end{equation}
with $u_{ns}$ and $u_{ms}$ being Wannier functions, give the
coupling between bands and sites. The specific value of $X_{s n,s'
m}$ is gave by the parameters of the material and the electric
field, but it is ``simulated'' by the RS. We note the rotator with a
linear $-i \frac{\partial}{\partial k}$ in the Hamiltonian is first
studied by Grempel \textit{et al} \cite{Grempel1982} and Berry
\cite{Berry1984}.

We assume $X_{s n,s' m}=X_{s n+n' ,s' m+n'}=X_{s0,s' m-n}$.
\begin{equation}
\begin{split}
H&=\sum_{s=1}^M\epsilon_s(k)|s\rangle \langle s|\\&-\sum_{s,s'=1}^M
eF \bigg(\sum_{n=-\infty}^\infty X_{s 0,s' n}e^{i n
k}\bigg)|s\rangle\langle s'| -i eF\frac{\partial}{\partial k},
\end{split}
\end{equation}
From Eq. $(49)$, we can also prove BO and SL based on the Floquet
theorem.

\section{\label{sec:level1}Conclusion and discussion}
In summary, we have given a proof of BO and SL based on the Floquet
theorem. We also give the exact solution of periodically driven two
level atom and two band BO in terms of Heun function.

The fully quantized field and two level atom interaction is more
interesting. But the present method completely fails because first,
after the second quantization, the Hamiltonian is not doubly
infinite; second, the off-diagonal matrix changes away from the
ground state \cite{Shirley1965}.

The original BO Hamiltonian Eq. $(31)$ can be approximated in the
$nk$ representation as in Eq. $(32)$ or in the site representation
as in Eq. $(11)$. In the $nk$ representation, we can prove BO and SL
from the Floquet theorem. In the site representation, we first
rewrite the same Hamiltonian in the rotator representation, then
again prove BO from the Floquet theorem. We do not know whether the
the two methods are equivalent. Although, the $nk$ representation
can be Fourier transformed into the site representation, it is
difficult to establish the equivalence between Eq. $(33)$ and
$(48)$. Both approximations may have their own merits in
applications. We note the multi-band approximation seems impossible
to remove the paradox of Zak \cite{Zak1968}.

Our method can not treat the most general BO problem with the
Hamiltonian Eq. $(31)$. But in a physical material, the Bloch
electron only occupies the lowest several bands. So we believe the
BO and SL are highly possible in reality.

Our most significant contribution in the paper is the equivalence
between the FS, BO and RS. We think the three are the same problem
with three ``faces''. The equivalence between FS and BO is
established by comparing the Floquet Hamiltonian Eq. $(29)$ and the
BO Hamiltonian Eq. $(30,32)$. The equivalence between BO and RS is
established by corresponding the site-band basis with the
rotation-spin basis. At last, the three share the same infinite
matrix structure.

The equivalence is shown as the following figure.
\begin{center}
\xymatrix { \boxed{\textbf{\txt{Floquet \\system}}}
\ar@{<->}[rr]^{\textbf{$nk$ representation}}
\ar@{<->}[dr]_{\textbf{Shirley, Sambe}}
                &  &    \boxed{ \textbf{\txt{Bloch \\oscillation}}} \ar@{<->}[dl]^{\textbf{site representation}} \\
                & \boxed{\textbf{\txt{Rotator \\ with spin}}}}
\end{center}

\fussy
\appendix
\section{Heun function}
The Heun function $H(a,q;\alpha,\beta,\gamma,\delta,z)$ is defined
as the solution of the following equations
\begin{equation}
\begin{split}
&\bigg(\frac{d^2}{dz^2}+(\frac{\gamma}{z}+\frac{\delta}{z-1}+\frac{\epsilon}{z-a})
\frac{d}{dz}+\frac{\alpha\beta z-q}{z(z-1)(z-a)}\bigg )\\
&\qquad \qquad H(a,q;\alpha,\beta,\gamma,\delta,z)=0,\\
&H(a,q;\alpha,\beta,\gamma,\delta,0)=1,\\
&\frac{dH(a,q;\alpha,\beta,\gamma,\delta,z)}{dz}|_{z=0}=\frac{q}{a\gamma},
\end{split}
\end{equation}
where $\epsilon=\alpha+\beta+1-\gamma-\delta$.
$H(a,q;\alpha,\beta,\gamma,\delta,z)$ is the second-order Fuchsian
equation with four regular singular points. One application of Heun
function is to the Calogero-Moser-Sutherland System
\cite{Takemura2003}.

The Heun confluent function $H_c(\alpha, \beta, \gamma, \delta,
\eta, z)$ is obtained from Heun function through a confluence
process. So $H_c(\alpha, \beta, \gamma, \delta, \eta, z)$ has two
regular singular points and one irregular one. It is the singular
points most important to our application. $H_c(\alpha, \beta,
\gamma, \delta, \eta, z)$ is defined as the solution of the
following equations
\begin{equation}
\begin{split}
&\bigg( z(z-1)\frac{d^2}{dz^2}+[\alpha z^2+(-\alpha+\beta+\gamma+2)z -\beta-1 ]\frac{d}{dz}+\\
&\frac{(\alpha(\beta+\gamma+2)+2\delta)z-\alpha(\beta+1)+\beta(\gamma+1)+\gamma+2\eta}{2}\bigg) \\
&\qquad \qquad \qquad H_c(\alpha, \beta, \gamma, \delta, \eta,z)=0,\\
&H_c(\alpha, \beta, \gamma, \delta, \eta, 0)=1,\\
&\frac{dH(\alpha, \beta, \gamma, \delta, \eta,
z)}{dz}|_{z=0}=\frac{\beta(-\alpha+\gamma+1)-\alpha+\gamma+2\eta}{2(\beta+1)}.
\end{split}
\end{equation}
\normalsize \nopagebreak The Heun doubly confluent function
$H_d(\alpha, \beta, \gamma, \delta, z)$ is obtained from Heun
function through two confluence process and has two irregular
singular points. $H_d(\alpha, \beta, \gamma, \delta, z)$ is defined
as the solution of the following equations
\begin{equation}
\begin{split}
&\bigg( (z-1)^3 (z+1)^3\frac{d^2}{dz^2}+(2z^5-\alpha
z^4-4z^3+2z+\alpha)\frac{d}{dz}\\
&+(\beta z^2+(2\alpha+\gamma)z+\delta) \bigg) H_d(\alpha, \beta,
\gamma, \delta, z)=0, \\
&H_d(\alpha, \beta, \gamma, \delta, 0)=1,\\
&\frac{H_d(\alpha, \beta, \gamma, \delta, z)}{dz}|_{z=0}=0.
\end{split}
\end{equation}
\nolinebreak
\begin{acknowledgments}
We would like to thank Professor Shaolong Wan for helpful
discussions. This work is supported by the National Natural Science
Foundation of China under Grant Numbers 10674125 and 10475070. S.-M.
Li is grateful to DFG for financial support during his stay in
Germany.
\end{acknowledgments}
\vspace*{5.0cm}
\newpage
\bibliography{Blochoscillation}
\end{document}